\documentclass[12pt]{article}
\usepackage{color,amsmath,amsfonts,amssymb,a4,epsfig}

\newcommand{\R}{{\mathbb{R}}}
\newcommand{\C}{{\mathbb{C}}}

\def\ha{\frac{1}{2}}

\def\pa{\partial}
\def\ra{\rightarrow}
\def\preuve{\begin{proof}}

\def\ge{\varepsilon}

\def\gf{\varphi}

\def\gl{\lambda}
\def\go{\omega}

\def\x{{\bf x}}
\def\xii{{\bf \xi}}

\def\OPD{~${\rm \Psi}$DO}

\def\nghbd{~neighbourhood~}
\newtheorem{defi}{Definition}

\newtheorem{lemm}{Lemma}

\newtheorem{theo}{Theorem}

\newtheorem{assumption}{Assumption}
\newenvironment{demo}{\noindent {\it Proof.--}
      \begin{quotation}\noindent}{\end{quotation}\hfill$\square $}

\begin{document}

\title{Mathematical models for passive imaging II: \\
 Effective Hamiltonians associated to surface waves.}
\author{Yves Colin de Verdi\`ere \footnote{Institut Fourier,
 Unit{\'e} mixte
 de recherche CNRS-UJF 5582,
 BP 74, 38402-Saint Martin d'H\`eres Cedex (France);
http://www-fourier.ujf-grenoble.fr/\~ ~ycolver/ 
}}


\maketitle

\begin{abstract}
In the present paper which follows our previous paper 
``Mathematical models for passive imaging I: general background'',
we discuss the case of surface waves in a medium 
which is stratified near its boundary  at some scale comparable 
to the wave length. We discuss how the propagation of such
waves is governed by effective Hamiltonians on the boundary.
The results are certainly not new, but we have been unable to find
a precise reference. They are very close to results in adiabatic
theory.
\end{abstract}

\section*{Introduction}
This paper is strongly related to the first part
\cite{YCdV2}. We will be more specific and discuss
the case of surface waves which are used in seismology
in order to image the earth crust.

We consider a medium $X$ with boundary $\pa X$
and assume that the medium is stratified near $\pa X$.
We
will discuss
how the linear  propagation of waves located near
the $\pa X$  is determined by
an effective Hamiltonian on $\pa X$.
It is interesting enough to remark that this Hamiltonian
is  no more a differential operator,
but only   a ``pseudo-differential operator'' (a \OPD) 
with a non-trivial dispersion relation (principal symbol). This  situation
is well known in physics as giving birth  to some kind of
wave guides.

A typical motivating situation is that of seismic surface waves propagating 
along the earth crust: it is well known that, at ``macroscopic scales'',
the earth crust is horizontally stratified such giving birth to
the so-called surface seismic waves which admit some more technical
names like ``Rayleigh'' or ``Love'' waves
(see \cite{Aki}). They play a crucial role in 
the passive imaging of the earth crust. Moreover, they are the most dangerous
in case of an earthquake.

From the general methods described in our paper \cite{YCdV2},
the field-field correlation allows to recover the dispersion 
relation of these surface waves from which we want to 
recover the transverse (vertical) structure of the crust.
After describing a general result which is close in spirit
to  the adiabatic Theorem in Quantum Mechanics, we will discuss
briefly what kind of inverse spectral problem we need to solve
at the end and we solve it under a physically realistic  monotonicity
assumption.

\section{The general setting}

We will work locally in
 $X=\{ ({\bf x},z)\in \R^{d-1}\times \R~|~ z\leq 0 \} $.
We will consider the very simple case of an acoustic wave equation
near the origine of $X$:
\begin{equation}\label{equ:ondes}
\left\{ \begin{array}{l}
u_{tt}-{\rm div}(n~{\rm grad}u)=0\\
u({\bf x},0)=0 
\end{array} \right.
\end{equation}
with 
\[ n({\bf x},z)=N({\bf x},z,\frac{z}{\ge})\]
and $N({\bf x},z,Z):\R^{d-1}\times \R_-\times \R _- \ra \R _+$
 a non negative  function which is independent of $Z$
for $Z\leq Z_0<0$.
There are 2 possible assumptions
on the regularity of $N$:
\begin{itemize}
\item \begin{assumption}\label{ass:A}
 $N$ is smooth
\end{assumption}
\item \begin{assumption}\label{ass:B}
$N$ is smooth as a function of $(\x, z)$ with values
into $L^\infty (\R_-)$ and 
$N(\x,z, Z)=N(\x,0,Z)+ O(z^\infty)$ \end{assumption}
\end{itemize}

In other words, the medium $X$ admits a quite irregular behaviour in
the vertical direction in a very small horizontal band $B_\ge \subset X$
 of width $\ge Z_0$.
We plan to see that Equation (\ref{equ:ondes}) admits, as $\ge \ra 0$,
asymptotic 
solutions of frequency of order
$\ge^{-1}$ located in $B_\ge $. Moreover these solutions are
determined by solving an effective pseudo-differential 
equation  on the boundary $\pa X= \R^{d-1}\times \{ 0 \}$.
\begin{assumption} \label{asump:slow}
We  will assume that 
\[0<  \inf _{ Z \leq 0}
N({\bf x},O,Z)<N({\bf x},0, -\infty  )~.\]
\end{assumption}
Physically, it means that the propagation speed at some points 
very close to   the boundary
is smaller than inside the medium. This kind of assumption is usually 
satisfied 
in seismology where the speed of elastic waves into surface sediments 
layers is smaller than the speed inside the rocks below the sediments.

\section{The main result}

\subsection{A Sturm-Liouville operator}

Let us consider, for each $(\x, \xi)\in T^\star \pa X $,
 the self-adjoint differential operator
$L_{\x ,\xii }$ on the half line $Z\leq 0$, with
Dirichlet boundary condition at $Z=0$, defined by:
\begin{equation} \label{equ:sturm}
L_{{\x,\xii }}v:=-\frac{d}{dZ}
(N({\bf x},0,Z)\frac{dv}{dZ}) + N({\bf x},0,Z)| \xii | ^2 v 
\end{equation}
In case of Assumption \ref{ass:B}, $L_{\x ,\xii }$
is defined in terms of quadratic forms.

The spectrum of $L_{\x ,\xii }$ consists of a finite discrete spectrum
and a continuous spectrum $[ N({\bf x},0,Z_0 )| \xii | ^2,+\infty [$.
Under  {\bf Assumption \ref{asump:slow}},  
$L_{{\x,\xii}}$ admits, for $\xii $ large enough, a non empty discrete spectrum
of simple eigenvalues
\[ \inf _Z N({\bf x},0,Z)| \xii | ^2 < \gl _1 ({\bf x,\xi})<
\cdots < \gl _j  ({\bf x,\xi})<
\cdots  < \gl _k  ({\bf x,\xi})< N({\bf x},0,Z_0 )| \xii | ^2~,\]
which depend smoothly of $(\x,\xii)$.
In order to see that, we can interpret  $ | \xii |^{-2}  L_{{\bf x,\xi}}$
as a semi-classical Schr\"odinger type operator with an effective
Planck constant $| \xii |^{-1}$ and a principal symbol
\[ p_{{\bf x }}(Z,\zeta)=
N({\bf x},0, Z)(\zeta ^2 + 1) \]
 which admits a 
well near $(Z, \zeta)=(0,0)$.
We should however take care of  the fact that the number 
$k$ depends on $({\bf x,\xi})$ and goes to $\infty $
as ${\bf \xi}$ does.

It leads to an interesting bifurcation problem which we will discuss
in Section~\ref{sec:bifurc}.

\subsection{WKB solutions of the  stationnary wave equation}

The goal of this section is to build the effective surface Hamiltonians
which describe the surface waves. Let us start with the:
\begin{lemm} \label{lemm:ansatz}
Let us consider the  operator $\hat{H}$ defined by:
\begin{equation}\label{equ:helm}
\hat{H}u:=-\ge ^2 {\rm div}(n~{\rm grad}u)
\end{equation}
acting on  functions on $X$  vanishing at $z=0$ (Dirichlet boundary 
conditions).

 Let us choose
$\gl  ({\bf x, \xi} )$ 
an eigenvalue of $L_{\bf x,\xi} $ depending smoothly of
$(\x, \xi) \in U $, where $U$
is a  bounded open set  of $T^\star \pa X $,
and $\gf  (\x,\xii, .) $ a normalized associated eigenfunction.
There exists:
\begin{itemize}
\item
 An asymptotic expansion
\[ \Phi _\ge=\sum _{m=0}^\infty \gf_{m}(\x,\xii,  \frac{z}{\ge})\ge^m \]
with $\gf_{0}=\gf $ and the  $\gf _{m}(\x,\xii, .)$'s 
are smoothly dependent of $(\x ,\xii)$ with values
  in the domain  of $L_{\x ,\xii }$. The
 $\gf_{m}~'s~(m\geq1)$ are unique if
they are assumed to be orthogonal to $\gf $.  
\item  A symbol
\[a_{\ge}(\x,{\bf \xi})= \sum _{m=0}^\infty a_{m}(\x,{\bf \xi})\ge^m \]
with $a_{0}=\gl $
\end{itemize}
such that we have the following identity of formal power series
in $\ge$:
\begin{equation} \label{equ:inter} \hat{H}\left(\Phi _{\ge}(\x, \xi, \frac{z}{\ge})
e^{i\langle \x | \xii \rangle /\ge  }\right)=
a_{\ge }(\x ,\xii )\Phi _{\ge}(\x, \xi, \frac{z}{\ge})
e^{i\langle \x | \xii \rangle/\ge  }~. \end{equation}

\end{lemm}
\begin{demo}
Expanding  $N$ by
Taylor formula, we get:
 $N(\x, \ge Z, Z)=\sum _{l=0}^\infty N_l(\x,Z)\ge ^l$
with
 \[N_l(\x, Z)=\frac{1}{l!}
\frac{\pa ^l N}{\pa z^l}(\x, O, Z) ~.\]

Under Assumption \ref{ass:A},
for $l\geq 1$, the $N_l$'s are smooth and  
 compactly supported in $Z$.
 Under Assumption  \ref{ass:B}, the $N_l$'s
 vanish for $l\geq 1$. 

We see that we are reduced to compute
\[ \Psi:= \hat{H}(e^{i\langle \x |\xi \rangle /\ge  }
\gf (\x, \frac{z}{\ge })) ~,\]
with $n(\x,z)=N(\x, z/\ge) $.
We find
\[ \Psi=e^{i\langle \x |\xi \rangle /\ge }\left( 
L_0\gf +\ge  L_1 \gf + \ge ^2 L_2 \gf \right) \]
with
$L_0=L_{\bf x,\xi}$, 
\[ L_1 \gf = -2in \langle \xi | \nabla _\x \gf \rangle 
-i \langle \xi | \nabla _\x n \rangle  \gf \]
\[ L_2 \gf = - \langle \nabla _\x n  | \nabla _\x \gf \rangle >
- n\Delta _\x \gf ~.\]

We start with $\gf _{0}=\gf  $ an eigenfunction of $L_{\x, \xi }$
with eigenvalue $\gl $.
By induction, we pick  all terms in $\ge ^m $ and get from
Equation (\ref{equ:inter}):
\begin{equation}\label{equ:eps_l}
 L_{\x, \xi} \gf_{m} =  \gl  \gf_{m}+ a_{m}\gf _0
 + R( \gf _{0},
\cdots , \gf _{m-1}) ~,\end{equation}
with 
\[  R( \gf _{0},
\cdots , \gf _{m-1})=\sum _{l=1}^{m-1}a_{m-l}\gf_l -
\sum_{l<m,~k+l+j=m}L_j^k \gf _l \]
where $L_j^k $ is given as $L_j$ with
$N$ replaced by $N_k$.

We first choose the unique $a_{m}$ so that 
 \[ a_{m}\gf_0 
 + R( \gf _{0},
\cdots , \gf _{m-1}) ~\]
is orthogonal to $\gf _0 $, then we can solve Equation (\ref{equ:eps_l}).
In order to show that   
 $\gf _{m}$ lies in the domain of  $L_{\x, \xi}$, it is enough
to show that $R( \gf _{0},
\cdots , \gf _{m-1})$ is   in $L^2 (Z)$.  

\end{demo}

\subsection{A short review  on micro-functions}

In order to help the reader, we give here a very short review
on micro-functions and  \OPD's.
Good references for more details could be \cite{Ma} or 
\cite{Gu-St}.

Let us first give the main  definitions:
\begin{defi} A family of functions (distributions) $f_\ge : X \ra \C$
is said to be {\rm admissible} if for any function
$\chi \in C_o^\infty (X)$, there exists some real number $s$ so that
the Sobolev norms $\| \chi f_\ge \|_s $ are  at most
of polynomial growth w.r. to $\ge ^{-1}$.
We will denote ${\cal A}(X)$ the vector space of such families.
\end{defi}

\begin{defi} The {\rm frequency set} or {\rm microsupport}, denoted
$WF(f_\ge)$, 
of an admissible family $f_\ge $ is the closed subset
of $T^\star X$ which is defined as follows in canonical
local coordinates
$(x, \xi)$:
\[ \begin{array}{l} (x_0,\xi_0)\notin  WF (f_\ge ) { ~if~and~only~if~}
\exists \chi \in C_o^\infty (X),~\chi (x_0)\ne 0,\\
~{\cal F}_\ge (\chi f_\ge )(\xi)= O(\ge ^\infty) 
{ ~for~}\xi { ~close~to~}\xi _0 ~.\end{array}\]
Here ${\cal F}_\ge u$ is the $\ge-$Fourier transform:
\[ {\cal F}_\ge u (\xi)=(2\pi \ge )^{-d/2}\int e^{-i\langle x|\xi \rangle 
/\ge } u(x) |dx | ~.\] 
\end{defi}

\begin{defi} If $X$ is a smooth manifold and 
$U$ is an open  set in $T^\star X $, 
$ {\cal M}(U)$ the space of {\rm microfunctions in $U$,}
is the quotient 
\[  {\cal M}(U):=
 {\cal A} (X) / \{ f_\ge | WF  (f_\ge )\cap U=\emptyset \}  ~.\]
 We will denote $f_\ge =O(\ge^\infty ) $ in $U$
the property 
\[ WF  (f_\ge)\cap U=\emptyset~.\] 
\end{defi}

\begin{defi}
If $a(x,\xi,\ge)$ is a suitable function, called a
{\rm symbol}, the {\rm pseudo-differential
operator} ($\Psi$DO) 
$A= {\rm Op}_\ge (a)$ is defined by:
\[ Au(x):=(2\pi \ge)^{-d}\int 
e^{i\langle x-y|\xi \rangle /\ge } a(x,\xi,\ge) u(y) |dy d\xi | ~.\]
\end{defi}

The main result is that $\Psi$DO's act on microfunctions:
\begin{theo} \label{theo:micros}
If $a(x,\xi,\ge)$ belongs to a suitable class of symbols
(for example, all derivatives of $a$ are uniformly bounded independently
of $\ge$), then 
\[ WF \left({\rm Op }_\ge (a) (f_\ge)\right)  \subset WF (f_\ge ) ~.\]

Moreover the action of ${\rm Op }_\ge (a)$
on ${\cal M}(U)$ depends only on the values of $a$ in some 
\nghbd of $\overline{U}$.
\end{theo}

\subsection{Reformulation of Lemma \ref{lemm:ansatz}
in terms of micro-functions}
From Lemma \ref{lemm:ansatz}, we get the
\begin{theo} Let us fix $(\x _0, \xi _0)\in T^\star \pa X$
 and assume that we have a smooth 
eigenvalue $\gl  (\x , \xii )$ of $L_{\x, \xii}$ defined
in some bounded \nghbd ~ $U' $ of $(\x _0, \xi _0)$.
Let us give $U $, an open \nghbd of $(\x _0, \xi _0)$,
 so that $\overline{U}\subset U'$
and $\chi \in C_o^\infty (U')$ which is $\equiv 1 $ on  $\overline{U}$.

The map:
 \[ J_{\ge}\left( e^{i\langle . |\xii \rangle /\ge  }\right)
(\x,z) := 
 \frac{1}{\sqrt{\ge} }\chi (\x,\xi)
 e^{i\langle \x |\xii \rangle /\ge  }\Phi  _{\ge }
\left(\x ,\xii ,\frac{z}{\ge }\right) \]
defines, by passing to the quotient,
  a    linear map
 $\overline{J_{\ge}} :{\cal M} (U)  \ra
{\cal M} (V) $, where 
\[ V:=\{ (\x,z;\xi,\zeta)\in T^\star X  ~|~(\x;\xi)\in U \}~,  \]
which 
satisfies
\begin{itemize}
\item $\overline{J_{\ge}}$ is microlocally unitary 
\item $\overline{J_{\ge}}$ is independent of the choice of $\chi $
\item $WF\left(\overline{J_{\ge}}(u)\right)\cap V \subset \{ z=0 \}$
\item  
$ \hat{H} \overline{J_{\ge}}  = \overline{J_{\ge}}
 \hat{H}_{ {\rm eff}}$,  
where $\hat{H}_{ {\rm eff}}$ is a \OPD ~ of full left symbol
$a_{\ge}(\x,\xi) $ and of principal symbol
$\gl  (\x, \xii )$
\end{itemize}
\end{theo}
\begin{demo}
The map $J_\ge$ is well defined because $\Phi _\ge (\x,\xi, Z) $
is well defined for $(\x,\xi)\in U' $. 
Let us explain first how $\overline{J_\ge }$ is defined.
We first can define $A_\ge $ by 
 \[ A_\ge  ={\rm Op}_\ge \left(\chi (\x,\xi )
 \Phi _\ge  (\x, \xi, . ) \right) ~,\]
as  a \OPD~ whose symbol takes values in $L^2 (\R^{-})$.
As such, by Theorem \ref{theo:micros},
 it is well defined from ${\cal M}(U)$
to ${\cal M}(U;L^2 (\R^{-})) $.
We now use the unitary map
$E_\ge :  {\cal M}(U;L^2 (\R^{-}))\ra  {\cal M}(V)$
which is defined by 
$E_\ge : f(x,.)\ra \ge ^{-\ha} f(\x,z/\ge )$
and define $J_\ge =D_\ge \circ A_\ge  $.

The rapid decay of $\Phi _\ge (\x,\xi,Z)$ w.r. to $Z$ implies the 
property  $WF\left(\overline{J_{\ge}}(u)\right)\cap V \subset \{ z=0 \}$.

The unitarity comes from the symbolic calculus:
$ A_\ge ^\star \circ D_\ge ^\star \circ D_\ge \circ A_\ge  $
is a scalar \OPD ~of principal symbol
$|\chi (\x,\xi)|^2 \int _{-\infty }^0
\ge ^{-1} |\phi (\x,\xi,\frac{z}{\ge})|^2 dz  $
which is $\equiv 1$ on $U$.

The interlacing property is easily checked on exponentials and
hence on microfunctions as a consequence of Lemma \ref{lemm:ansatz}.

\end{demo}

\subsection{Speeds of propagation}

To any dispersion relation $ K (\x, \xii )= \go ^2$ is associated 
a speed of propagation (the so-called ``group-speed'')
defined by 
\[ {\bf v}:=\frac{d \x }{dt}={\pa _\xi K}/ 2 \go ~.\]

For the acoustic equation, we have 
$n(\x, z)|  \xi | ^2= \go ^2$ and hence
${\bf v}= n \xi / \go $
and $v:=|{\bf v}|= \sqrt{n} $.
We see the known result that the speed is independent of the frequency.

Let us compute the speed for
$\gl  (\x, \xii)=\go ^2$:
we have 
\[ \pa _\xi \gl  (x,\xi ) = 2  \left( \int _{-\infty}^0 N(\x , 0, Z)
 \gf  (Z)^2 dZ\right)\xii
\]
and hence
\[ v =\frac{|  \xii  |}{\go}
  \left( \int_{-\infty}^0 N(\x , 0, Z) \gf  (Z)^2 dZ \right)\]
and using 
\[ \int_{-\infty}^0 N(\x , 0, Z)|\xi| ^2  \gf  (Z)^2 dZ < \gl  =\go ^2 \]
we get
\[   v < \frac{\go}{|\xi|}< \sqrt{n(\x,0, Z_0)} ~.\]
{\it The speed of the surface waves is less than the speed of the body waves.}

\section{An inverse spectral problem}

Recovering the vertical structure of earth crust from
the dispersion relation of surface waves needs to solve the following inverse
problem:

{\bf Let us assume that we know the discrete spectrum
of $L_{\x, \xi} $ at fixed $\x$ for all $\xi$'s, can we recover
the vertical profile  $N(\x,0, Z)$?} 

Let us show the
\begin{theo}
If $N(\x,0,Z)$ is a {\rm decreasing
function} of $Z$ from $N(\x, 0, Z_0)$ to $N(\x, 0, 0)$,
the asymptotics of the discrete spectra $\gl _j (\x, \xi),~1\leq j
 \leq k(\x, \xi) $ as  $\xi \ra \infty $
 determine the functions $N(\x,0,Z).$
\end{theo}
\begin{demo} This is a consequence of Weyl's asymptotics:
for $E< N(\x, 0, Z_0)$, let us introduce
\[ N(\x, \xi, E)=\# \{ \gl _j (\x, \xi) \leq E |\xi| ^2 \} ~.\]
We have the  following (Weyl type) asymptotics as $\xi \ra \infty $:
\[ N(\x, \xi, E) \sim \left( \frac{|\xi |}{2\pi }\right)^\ha
{\rm Area}\left( \{ N(\x, 0, Z)(1+\zeta ^2) \leq E \} \right) ~.\]
It implies that, from the asymptotics of $ \gl _j (\x, \xi)$,
we can recover
\[ V(E)=\int _{z(E)}^0 \sqrt{\frac{E}{N(Z)}-1}~ dZ ~,\]
with $N(z(E))=E $.
We can rewrite:
\[ V(E)=\int _{N_0}^E \sqrt{E-u}~\frac{z'(u)~du}{\sqrt{u}} \]
and introducing 
\[ Kf(E):=\int _{N_0}^E \sqrt{E-u}~ f(u)~ du \]
we get 
\[ \frac{d^3}{dE ^3} ( (K\circ K) f) (E)=2 f(E) \]
which allows to recover $z'(E)/\sqrt{E}$ as the third order derivative
of $K(V)(E)  $.

\end{demo}

It is a quite interesting problem to try to extend the 
previous Theorem without the assumption that $N$ is decreasing.
In this case Weyl formula is no more  enough,
 but more refined trace formulae like Gutzwiller
trace formula may be usefull to give a general answer.

\section{Bifurcations} \label{sec:bifurc}

What happens when the eigenvalue $\gl _k (\x, \xi)$ goes to 
the bottom of the continuous spectrum.
It is an interesting place when there should be a mode conversion
from body waves into surface waves or conversely.
 It could be interesting from the point
of view of the study of earthquakes whose source is deep: how do they give 
birth to surface waves?
It could also be interesting in order to study 
resonances created by surface waves into a bassin.
We plan to study these (difficult?) questions in a future work.


\bibliographystyle{plain}

\end{document}